      \documentstyle[12pt]{article}
\def\maketitle{\par
 \begingroup
 \def\thefootnote{\fnsymbol{footnote}}
 \def\@makefnmark{\mbox{$^\@thefnmark$}}
 \@maketitle
 \@thanks
 \endgroup
 \setcounter{footnote}{0}
 \let\maketitle\relax
 \let\@maketitle\relax
 \gdef\@thanks{}\gdef\@author{}\gdef\@title{}\let\thanks\relax}
\def\@maketitle{\vspace*{0.9cm}
{\hsize\textwidth
 \linewidth\hsize \centering
 {\normalsize \bf \@title \par} \vskip 1.0cm  {\normalsize  \@author \par}}}

\def\thefootnote{\mbox{\noindent$\fnsymbol{footnote}$}}
    \long\def\@makefntext#1{\noindent$^{\@thefnmark}$#1}

\def\citenum#1{{\def\@cite##1##2{##1}\cite{#1}}}
\def\citea#1{\@cite{#1}{}}

\def\lsim{\mathrel{\rlap{\lower4pt\hbox{\hskip1pt$\sim$}}
    \raise1pt\hbox{$<$}}}         
\def\gsim{\mathrel{\rlap{\lower4pt\hbox{\hskip1pt$\sim$}}
    \raise1pt\hbox{$>$}}}         

\def\beq{\begin{equation}}
\def\endeq{\end{equation}}
\def\arr{\begin{eqnarray}}
\def\endarr{\end{eqnarray}}

\makeindex

\begin{document}
\phantom{.}\hspace{7.5cm}{\bf \Large KFA-IKP(Th)-1993-28}\\
\phantom{.}\hspace{9.5cm}{\sl 4 November 1993}\vspace{2.0cm}\\
\begin{center}
{\bf \huge COLOR TRANSPARENCY \\
AFTER THE NE18 AND E665 EXPERIMENTS$^{*}$\vspace{1cm}\\}
{\bf \large underline{N.~N.~Nikolaev}$^{1,2)}$
and B.~G.~Zakharov$^{1}$\smallskip\\ }
{\em
$^{1)}$L.D.Landau Institute for Theoretical Physics,
Kosygina 2, V-334 Moscow, Russia\\
$^{2)}$IKP(Theorie), KFA  J\"ulich GmbH.,
D-52425 J\"ulich, Germany
\vspace{1.0cm}\\}
{\bf ABSTRACT}
\setlength{\baselineskip}{2.6ex}
\end{center}
{
The 1993 witnessed two major news in the field of color
transparency (CT):
i) no effect
of CT was seen in the SLAC NE18 experiment on
$A(e,e'p)$ scattering at virtualities of the exchanged photon
 $Q^{2} \lsim 7$ GeV$^{2}$, ii)
 strong signal of CT was observed
in the Fermilab E665 experiment on exclusive $\rho^{0}$-meson
production in deep inelastic scattering
in the same range of $Q^{2}$. Both are good news and rule in
CT, since this striking
difference in the
onset of CT for two reactions was predicted theoretically,
and we review the theory of both reactions and
the origin of this difference. We discuss an
importance of final state interaction effects for the
theoretical interpretation of the NE18 data. We comment on
possible CT effects in the $^{4}He(e,e'p)$ interactions at CEBAF.
} \bigskip\\
\begin{center}
{\bf \large $^{*)}$~Invited talk at the Workshop on\\
Electrom Nucleus Scattering\\
Elba International Physics Center\\
Marciana Marina, July 5-10, 1994}
\pagebreak\\
\end{center}
{\bf 1.~ Introduction}\smallskip\\
In QCD quark configurations with small transverse size
$\vec{r}$ have small interaction cross section [1]. Such small-sized
configurations are expected to emerge from exclusive
hard scattering vertices which are selective to $r\sim 1/Q$, where
$Q$ is the momentum transfer; the quasielastic $A(e,e'p)$ and
$A(p,p'p)$ scattering being the typical candidate
reactions [2]. The signal of CT in production on nuclei is
vanishing strength of final state interaction (FSI) at
$Q\rightarrow \infty$. The $A(e,e'p)$ reaction
on the $D$, $C$, $Fe$ and $Au$ targets was studied by the SLAC NE18
collaboration with the negative result: no CT effects
are seen at $Q^{2} \leq 7$ GeV$^{2}$ [3].
The negative result of the NE18 experiment dampened
expectations of precocious CT at low $Q^{2}$, claimed in
numerous papers based on the theoretically inconsistent
approach to CT (for the review
see [6,7]).
Such a slow
onset of CT in the $A(e,e'p)$ scattering was predicted in our
papers [4,5] (for the review see [6,7]) and in the second
part of this talk we summarize deep reasons for this slow onset.
As a matter of fact, the NE18 results do perfectly
confirm the correct theory,
CT is alive and well, and we can joyfully recite
Mark Twain's telegram to the Associated Press:
"The reports of my death were an exaggeration".

The parallel development was a theory of CT in
(virtual) photoproduction of vector mesons
$\gamma^{*}N\rightarrow VN$
in deep inelastic scattering [8-13].
In our work [11] we have predicted
fast onset of CT in the $\gamma^{*}N\rightarrow VN$
production.
This
prediction was confirmed by the Fermilab E665 experiment [14],
which found the solid evidence for CT in precisely the same
range of $Q^{2}$ as explored in the NE18 experiment.

In the first part of this talk we review our work on CT
in the
vector meson production [6-13], the second part is
based on works on $A(e,e'p)$ scattering by the
ITEP-J\"ulich-Krakow-Landau collaboration [4-7,15-19].
\bigskip\\
\noindent
{\bf 2.~ CT in exclusive production of vector mesons}
\medskip\\
{\sl 2.1. The lightcone approach to virtual photoproduction}
\medskip\\
The reaction $\gamma^{*}p\rightarrow Vp$   ($V=\rho^{0},\phi^{0},
J/\Psi$,...)
is an ideal laboratory for testing
CT ideas. It is
a hard scattering process in which the transverse
size $r_{Q}$ of quark configurations which dominate the production
amplitude is under the good control [11],
\beq
r_{Q} = {2 \over \sqrt{m_{V}^{2}+Q^{2}}} \, ,
\label{eq:2.1.1}
\endeq
In the E665 experiment [14] the produced $\rho^{0}$ has an energy
$\nu \sim 200$ GeV, and the reaction mechanism
greatly simplifies: The photon fluctuates into the $q\bar{q}$ pair at
a large distance (the coherence length)
$
l_{c}={{2\nu}/(Q^{2}+m_{V}^{2})}$
in front of the target nucleon (nucleus). After interaction  the
$q\bar{q}$ pair recombines into the vector meson $V$ with the
recombination (formation) length
$
l_{f}={\nu/ m_{V}\Delta m}, $
where $\Delta m $ is the typical level splitting in the
quarkonium. At high energy $\nu$ both $l_{c}$ and $l_{f}$  greatly
exceed the radius $R_{N}$ ($R_{A}$) of the target nucleon (nucleus),
and the transverse
size $\vec{r}$ of the $q\bar{q}$ pair and the longitudinal
momentum partition $z$ and $(1-z)$ between the quark and antiquark
of the pair do not change during the interaction with the target.
This enables one to introduce the lightcone wave function
$\Psi_{\gamma^{*}}(r,z)$ of the $q\bar{q}$ fluctuation
[12]. The color-singlet  $q\bar{q}$ pair
interacts with the target nucleon with the cross
section [12,20]
\beq
\sigma(r) ={\pi^{2}\over 3}r^{2}
{\cal F}(\nu,r)
\label{eq:2.1.2}
\endeq
where ${\cal F}(\nu,r)$ is related at small $r$ to the gluon
structure function of the proton, ${\cal F}(\nu,r)=
\alpha_{S}(r)xg(x,Q_{r}^{2})$, evaluated at
$Q_{r}^{2} \sim    1/r^{2}$ and at the Bjorken variable
$x\approx (Q_{r}^{2}+m_{V}^{2})/2m_{N}\nu$,
and
$\alpha_{S}(r)$ is the running QCD coupling. This factor
${\cal F}(\nu,r)$
takes into account the effect of higher $q\bar{q}g_{1}...g_{n}$
Fock components in the lightcone wave function of the photon
and the vector meson [20].
It is a smooth function of $r$ compared to $r^{2}$ in
Eq.~(\ref{eq:2.1.2}). The cross section $\sigma(r)$
vanishes  at
$r\rightarrow 0$ ant its $r$ dependence quantifies CT property
in QCD.

Because of the frozen $\vec{r}$, the
amplitude of the forward photoproduction
$\gamma^{*}N\rightarrow VN$ can be cast [8,11]
in the quantum-mechanical form [$\vec{q}$ is the momentum transfer]
\beq
M(VN,\vec{q}=0)=\langle V|\sigma(r)|\gamma^{*}\rangle
=\int_{0}^{1} dz \int d^{2}\vec{r}\,\sigma(r)\Psi_{V}(r,z)^{*}
\Psi_{\gamma^{*}}(r,z)
\label{eq:2.1.3}
\endeq
The most important feature of
$\Psi_{\gamma^{*}}(r,z)$ is an exponential decrease
at large size [12]:
\beq
\Psi_{\gamma^{*}}(r,z)
\propto \exp(-\varepsilon r)
\label{eq:2.1.4}
\endeq
where
\beq
\varepsilon^{2} = m_{q}^{2}+z(1-z)Q^{2} \approx
{1\over 4}(m_{V}^{2} + 4z(1-z)Q^{2})
\label{eq:2.1.5}
\endeq
In the nonrelativistic quarkonium $z \approx 1/2$,
$\Psi_{\gamma^{*}}(r,z)$ is concentarted at $r \lsim r_{Q}$ and
the wave function (\ref{eq:2.1.4}) [12] and
Eq.~(\ref{eq:2.1.1}) for $r_{Q}$ [11] fulfill the dream [2]
of having the
well specified control of the size of quark configurations
important in the hard scattering process.

The generalization to nuclear targets is straightforward [8-11]:
The nuclear transparency  for the incoherent production
equals
\beq
T_{A}={\sigma_{A} \over A\sigma_{p}}={1\over A}
\int d^{2}\vec{b} T(b)
{\langle V |\sigma(r)
\exp\left[-{1\over 2} \sigma(r)T(b)\right] |\gamma^{*}
\rangle^{2} \over
\langle V|\sigma(r)|\gamma^{*}\rangle^{2} } =
1-\Sigma_{V} {1\over A}\int d^{2}\vec{b}T(b)^{2} +.....
\label{eq:2.1.6}
\endeq
where $T(b)=\int dz n_{A}(b,z)$ is the optical thickness
of a nucleus at the impact parameter b.
The total
cross section of the coherent production equals
\arr
\sigma_{coh}(VA) =
4\int d^{2}\vec{b}
\left|\langle V |1-
\exp\left[-{1\over 2} \sigma(r)T(b)\right] |\gamma^{*     }
\rangle\right|^{2}\nonumber\\ =
\left|M(VN,\vec{q}=0)\right|^{2}\int d^{2}\vec{b}
T(b)^{2}\left[1-
{1\over 2}\Sigma_{V}T(b) +... \right]
\label{eq:2.1.7}
\endarr
We have explicitly shown the leading terms of FSI.
The strength of FSI is measured by the observable [4,9]
\beq
\Sigma_{V}={\langle V|\sigma(r)^{2}|\gamma^{*}\rangle
\over \langle V|\sigma(r)|\gamma^{*}\rangle } \,.
\label{eq:2.1.8}
\end{equation}
\noindent
{\sl 2.2. How $\sigma(r)$ is probed in the virtual photoproduction}
\medskip\\
The wave function of the vector meson is smooth at small $r$.
Because of CT property (\ref{eq:2.1.2}) the  integrand
of (\ref{eq:2.1.3}) is $\propto r^{3}\exp(-{r\over r_{Q}})$ and
the amplitude of (virtual)
photoproduction will be dominated by contribution
from size [9-11]
\beq
r_{S} \sim 3r_{Q}
\label{eq:2.2.1}
\endeq
which falls in into the perturbative QCD domain $r_{S} \ll R_{V}$
at a sufficiently large $Q^{2}$. In this domain for production of
transversely polarized vector mesons by the transversely polarized
photons $\gamma_{T}^{*} N \rightarrow V_{T}N$, Eq.~(\ref{eq:2.1.3})
gives an estimate [11]
\beq
M_{T}(VN,\vec{q}=0)
\propto {r_{S}^{2} \over R_{V}^{3/2}}\sigma(r_{S})
\propto {1\over (Q^{2}+m_{V}^{2})^{2}}
{\cal F}(\nu,r_{S}) \, .
\label{eq:2.2.2}
\endeq
The $(m_{V}^{2}+Q^{2})^{-2}$ behavior of the
amplitude (\ref{eq:2.2.2})
is different from the VDM prediction [21,22], in good
agreement with
the experiment [14,23].
This difference
comes from the factor $\sigma(r_{S})\sim    1/(Q^{2}+m_{V}^{2})$
in (\ref{eq:2.1.3}) which
emphasizes a relevance of CT property of
$\sigma(r)$ to the total production rate.
Relativistic effects in the wave function of vector mesons
somewhat slow down the rapid decrease (\ref{eq:2.2.2}) [11,24]

The integrand of
$\langle V|\sigma(r)^{2}|\gamma^{*}\rangle$ in the strength
of FSI $\Sigma_{V}$ is
$\sim r^{5}\exp(-{r\over r_{Q}})$ and
$\langle V|\sigma(r)^{2}|\gamma^{*}\rangle$ will
be dominated by
\beq
r\sim r_{FSI}=(4-5)r_{Q} \, ,
\label{eq:2.2.3}
\endeq
which gives an estimate [11]
\beq
\Sigma_{V} \approx \sigma(r_{FSI})
\label{eq:2.2.4}
\endeq
CT and/or weak FSI set in when $\Sigma_{V} \ll
\sigma_{tot}(VN) \approx \sigma(R_{V})$.
In this regime of CT
$\Sigma_{V}$ is insensitive to
the wave function of the vector meson, so that
predictions of FSI effects are model independent.
\noindent
\medskip\\
{\sl 2.3. The E665 experiment: the decisive proof of CT}
\medskip\\
In Fig.1 we show our predictions [11] for
nuclear transparency for the incoherent
cross section as a function of $Q^{2}$.
Nuclear attenuation
is very strong at small $Q^{2}$ and gradually decreases with
$Q^{2}$. This rise of nuclear transparency $T_{A}$ with $Q^{2}$
is particularly dramatic for the heavy
nuclei ($Ca,\, Pb$), and leaves no doubts the E665 observed the
onset of CT.
In Fig.~2 we present our predictions [11] for the $Q^{2}$ dependence
of nuclear transparency for the forward coherent production on nuclei
$
T_{A}^{(coh)} = {d\sigma_{A}^{(coh)}/A^{2} d\sigma_{N}}
|_{\vec{q}\,^{2}=0}$ .
We predict a rise of $T_{A}^{(coh)}$ with $Q^{2}$ towards
$T_{A}^{(coh)}=1$ for the complete CT.
In Fig.~3 we present our predictions [11] for the $Q^{2}$ dependence
of the coherent production cross section relative
to the cross section for the carbon nucleus.
For the regime of complete CT and/or vanishing FSI ($R_{ch}(A)$ is
the charge radius of a nucleus),
\beq
R_{coh}^{(CT)}(A/C)= {12 \sigma_{A} \over A\sigma_{C}}
\approx {AR_{ch}(C)^{2} \over 12R_{ch}(A)^{2}}\, ,
\label{eq:2.3.1}
\endeq
which gives $R_{coh}^{(CT)}(Ca/C)= 1.56$ and
$R_{coh}^{(CT)}(Pb/C) = 3.25$. The observed growth of the
$Pb/C$ ratio gives
a solid evidence for the onset of CT effects.

The (approximate) $A^{\alpha}$ parametrization is a convenient
short-hand representation of the $A$-dependence of nuclear
cross sections, although the so-defined exponent $\alpha$ slightly
depends on
the range of the mass number $A$ used in the fit.
Then, Eq.~(\ref{eq:2.1.6}) predicts that $\alpha_{inc}(Q^{2})$ tends
to 1 from below, as $Q^{2}$ increases. In the limit of vanishing
FSI Eq.~(\ref{eq:2.1.7}) predicts
$\sigma_{coh} \sim    A^{4/3}$, so
that $\alpha_{coh}(Q^{2})$ tends to $\approx {4\over 3}$
from below as $Q^{2}$ increases
(more accurate analysis shows that the no-FSI cross
section in the $C-Pb$ range of nuclei has $\alpha_{coh} \sim 1.39$).
The agreement between the theory and the E665 fits is good (Fig.4).
Both the $\alpha_{coh}(Q^{2})$ and
$\alpha_{inc}(Q^{2})$ rise with $Q^{2}$, which is still another
way of stating that the E665 data confirm the onset of CT.
\noindent
\medskip\\
{\sl 2.4. The precocious CT?}
\medskip\\
Remarkably,
the large numerical factor $\approx (4-5)$ in the  r.h.s.
of Eq.~(\ref{eq:2.2.2}) comes from CT property of $\sigma(r)$,
and the very CT property of the production mechanism
predicts a belated onset of the CT effect in
nuclear attenuation, which
requires $r_{Q} \ll {1\over 5}R_{V}$.
Although at the highest
$Q^{2} \sim 10$ GeV$^{2}$ of the E665 experiment
Eq.~(\ref{eq:2.1.1}) gives $r_{Q} \approx 0.13$ f ,
Eq.~(\ref{eq:2.2.3}) shows that
$r_{FSI} \approx 0.5$ f and FSI is still substantial (see also
section 2.6).
The large value of $r_{FSI}$ also
shows that the relativistic effects in vector mesons
are not yet important
in the nuclear attenuation calculations.
\noindent
\medskip\\
{\sl 2.5 Polarization dependence of the vector meson production}
\medskip\\
The free-nucleon reaction mechanism changes drastically with $Q^{2}$.
{}From the essentially kinematical considerations one
finds a dominance of the longitudinal cross section
at large $Q^{2}$ [22]
\beq
{\sigma_{L} \over \sigma_{T}} \approx {Q^{2} \over m_{V}^{2}} \, ,
\label{eq:2.5.1}
\endeq
in good agreement with the E665 [14]
and NMC [23] data. The
relationship (\ref{eq:2.5.1}) holds for the nonrelativistic
quarkonium and the relativistic effects slow down the rise
(\ref{eq:2.5.1}) [11,24].
Even with allowance for the relativistic corrections,
FSI effects in $\sigma_{L}$
and $\sigma_{T}$ differ little, and we predict approximately
$A$-independent polarization
density matrix of the produced $\rho^{0}$ mesons [11].
The E665 data [14] confirm this prediction.
\noindent
\medskip\\
{\sl 2.6.  Scaling properties of CT effects}
\medskip\\
We predict that in the regime of CT
$1-T_{A}$ scales with $r_{Q}^{2}$, {\sl i.e.,} with
$(Q^{2}+m_{V}^{2})$ [11]
\beq
1-Tr_{A} \propto
{A\over R_{ch}(A)^{2}}r_{S}^{2}{\cal F}(\nu,r_{S}) \sim
{A\over R_{ch}(A)^{2}}{1\over Q^{2}+m_{V}^{2}}
\label{eq:2.6.1}
\endeq
(At a small and moderate $Q^{2}$, when $r_{FSI}\sim R_{V}$, the
observable $\Sigma_{V'}$ for the production of the radially
excited vector mesons $V'$
is extremely sensitive to the nodal
structure of the wave function of the $V'$,
which may lead to $\Sigma_{V'} <0$ and
to the antishadowing phenomenon [8-11].)
The scaling law (\ref{eq:2.6.1}) predicts that
at $Q^{2} \approx 9$ GeV$^{2}$
the nuclear attenuation
for the incoherent $\rho^{0}$ production must be the same
as for the incoherent real ($Q^{2}=0$)
photoproduction of the $J/\Psi$.
For the $\rho^{0}$ production
at $\langle Q^{2} \rangle =7$ GeV$^{2}$ the
E665 experiment gives $T_{Pb}/T_{C} = 0.6 \pm 0.25$.
This can be
compared with the NMC result $T_{Sn}/T_{C}= 0.7 \pm 0.1$ for
the real photoproduction of $J/\Psi$ in the similar energy range [25].
This substantial departure from the complete CT,
$T_{A}/T_{C}=1$, confirms our conclusion
that even at $Q^{2}\sim 10$ GeV$^{2}$
the FSI is controlled by a large $r_{FSI} \sim 0.5$ f and is not
yet vanishing.
Similar scaling law holds for the coherent production of
the $\rho^{0}$ and the $J/\Psi$.
In the regime of complete CT
Eq.~(\ref{eq:2.1.6}) gives [10,11]
$R_{coh}(Sn/C)=2.76$, $R_{coh}(Fe/Be)=2.82$, $R_{coh}(Pb/Be)=
4.79$. The experimental data on the real photoproduction
of $J/\Psi$, which were sucsessfully described within the discussed
framework  [8,10],
give a solid evidence for substantial FSI:
$R_{coh}(Sn/C)=2.15\pm 0.10$ in the NMC experiment [25] and
$R_{coh}(Fe/Be)=2.28\pm 0.32$, $R_{coh}(Pb/Be)=3.47 \pm 0.50$
in the Fermilab E691 experiment [26]. In all cases the observed
$\sim 30\%$
departure
of the observed ratios for the $J/\Psi$ from predictions for the
complete CT
is of the same magnitude as in the highest $Q^{2}$
bin of the E665 data on the $\rho^{0}$ production (Fig.3).
Higher precision data on the $J/\Psi$ and $\rho^{0}$ production
at higher $Q^{2}$ would be very interesting for tests of our scaling
law (\ref{eq:2.6.1}).
\noindent
\medskip\\
{\sl 2.7. Measuring the wavefunction of vector mesons}
\medskip \\
The relativistic effects slow down the
rapid decrease (\ref{eq:2.2.2})
of $M_{T}$ with $Q^{2}$ [11,24]. Because of
CT property of $\sigma(r)$ large values of $r$ are favored
in the integrand of the production amplitude (\ref{eq:2.1.3}).
The photon
wavefunction $\Psi_{\gamma^{*}}(r,z)$ admits large $r$ if
$\varepsilon$ Eq.~(\ref{eq:2.1.5}) is small, {\sl i.e.,} if either
$z$ or $(1-z)$ is small. Although such asymmetric pairs in the
vector meson correspond to the large
longitudinal momenta of quarks, which suppresses the
wavefunction of the vector meson, at very large $Q^{2}$
the production amplitude will be dominated by the asymmetric
quark configurations, rather than by the symmetric
nonrelativistic configurations $z\sim 1/2$.
Because $\Psi_{\gamma^{*}}(r,z)$ and $\sigma(r)$ are
theoretically well understood [12,20],
the virtual
photoproduction offers a unique opportunity of scanning and
measuring the wave function of vector mesons [11].
\pagebreak\\
\noindent
{\bf 3.~ CT in $A(e,e'p)$ scattering}
\medskip\\
{\sl 3.1. Nuclear transparency, FSI and spectral function}
\medskip\\
We start with formulation of predictions of the standard multiple
scattering theory. The quantity measured in $A(e,e'p)$ scattering
is the spectral function $S(E_{m},\vec{p}_{m})$. If the measured
cross section is integrated over the sufficiently broad range of
the missing energy $E_{m}$, then in the plane-wave impulse
approximation (PWIA) one measures the single-particle momentum
distribution related to the one-body
nuclear density matrix $\rho_{1}(\vec{r},\vec{r}\,')$ as
\arr
d\sigma_{A}
\propto
 n_{F}(\vec{p}_{m})=
{1\over Z}\int dE_{m}S_{PWIA}(E_{m},\vec{p}_{m})=
\int d\vec{r}\,'d\vec{r} \,
\rho_{1}(\vec{r},\vec{r}\,')
\exp[i\vec{p}_{m}(\vec{r}\,'-\vec{r})]
\label{eq:3.1.1}
\endarr
The FSI modifies Eq.~(\ref{eq:3.1.1}) [17,18]:
\arr
d\sigma_{A} \propto
{1\over Z}\int dE_{m}S(E_{m},\vec{p}_{m})=
\int d\vec{r}\,'d\vec{r} \,
\rho_{1}(\vec{r},\vec{r}\,')
\exp[i\vec{p}_{m}(\vec{r}\,'-\vec{r})]
\cdot\exp\left[ t(\vec{b},max(z,z')) \xi(\vec{\Delta}) \right]
\nonumber \\
\cdot \exp\left[
-{1\over 2}(1-i\alpha_{pN})\sigma_{tot}(pN)
t(\vec{b},z)
-{1\over 2}(1+i\alpha_{pN})\sigma_{tot}(pN)
t(\vec{b}',z') \right] \,,~~~~~~~~~~~~~~~~
\label{eq:3.1.2}
\endarr
where $\vec{r}=(\vec{b},z),\,\vec{r}\,'=(\vec{b}',z')$,
$\vec{\Delta}=\vec{r}-\vec{r}\,'$, $\alpha_{pN}$
denotes the ratio of the real to imaginary parts of the forward
proton-nucleon scattering amplitude and
\beq
\xi(\vec{\Delta}) =
\int d^2\vec{q} \;\
\frac{d\sigma_{el}(pN)}{d^2\vec{q}} \;\
\exp(i \vec{q}\vec{\Delta}) \, .
\label{eq:3.1.3}
\endeq
FSI leads to the two important effects:
Firstly, as a result
of the phase factor in the integrand of
(\ref{eq:3.1.2}) which is of the form
$\exp(i\sigma_{tot}(pN)\alpha_{pN}n_{A}(b,z)(z-z'))$
the spectral function  $S(k_{\perp},k_{z})$ is probed
at a shifted value of the longitudinal momentum with [18]
\beq
k_{z}- p_{m,z}=  \Delta p_{m,z}
\sim
\sigma_{tot}(pN)\alpha_{pN}n_{A} \approx
\alpha_{pN}\cdot 70\, {\rm (MeV/c)} \, .
\label{eq:3.1.4}
\endeq
In the kinematical range of the
NE18 experiment $\alpha_{pN} \sim -0.5$ [27] and $\Delta p_{m,z}
\sim -35$ MeV$/c$ is quite large.
Secondly, the factor
$\exp\left[ t(\vec{b},z) \xi(\vec{\Delta}) \right]$ in
Eq.~(\ref{eq:3.1.2}) leads to a
broadening of the $p_{\perp}$ distribution. The
multiple elastic-rescattering expansion
for the $(E_{m},p_{m,z})$-integrated spectral
function reads [17,18,6]
\beq
f{_A}(\vec{p}_{\perp}) = {1\over Z}\int dE_{m}\,dp_{m,z}\,\,
S(E_{m},p_{m,z},p_{\perp})=
\sum_{\nu = 0}^{\infty}
W^{(\nu)}\,n^{(\nu)}(\vec{p}_{\perp}) \, ,
\label{eq:3.1.5}
\endeq
where
\beq
W^{(\nu)} = \frac{1}{A} \int dz d^2\vec{b} \;\ n_A(\vec{b},z)\,
\exp \left[-\sigma_{tot}(pN) t(\vec{b},z) \right]
{[t(\vec{b},z)
\sigma_{el}(pN)]^{\nu}\over \nu! }
\label{eq:3.1.6}
\endeq
gives the probability of having $\nu$ elastic rescatterings.
These probabilities $W^{(\nu)}$ are normalized as
\beq
\sum_{\nu=0} W^{(\nu)}=T_{A}
 = \frac{1}{A} \int dz d^2\vec{b} \;\ n_A(\vec{b},z)
\exp \left[-\sigma_{in}(pN) t(\vec{b},z) \right]\, .
\label{eq:3.1.7}
\endeq
Here $t(b,z)=\int_{-\infty}^{z} dz'\,n_{A}(b,z')$,~~
$T(b)=t(b,\infty)$ and  the $p_{\perp}$-distribution in the
$\nu$-fold rescattering $n^{(\nu)}(\vec{p}_{\perp})$ equals
$
n^{(\nu)}(\vec{p}_{\perp}) =
\int d^2 \vec{s}  \;\ (B/\nu\pi)
\exp \left( - B s^2/\nu \right)
n_{F}(\vec{p}_{\perp} - \vec{s})$,
where $B$ denotes the diffraction slope for elastic $pN$
scattering.
If the spectral function (\ref{eq:3.1.5}) is
integrated over the entire range of $p_{\perp}$,
then $T_{A}$ as given by Eq.~(\ref{eq:3.1.7}), {\sl i.e.,} the
attenuation is given by $\sigma_{in}(pN)$ [5].
On the other hand, the forward peak of $f_{A}(p_{\perp})$ at
$\vec{p}_{\perp}=0$  is dominated by
$W^{(0)}$, which is also given by
Eq.~(\ref{eq:3.1.7}) but with $\sigma_{in}(pN)$ substituted by
$\sigma_{tot}(pN)$.
\noindent
\medskip\\
{\sl 3.2.  FSI in the $d(e,e'p)$ scattering}
\medskip\\
The momentum distribution $f_d(\vec p)$ of the observed protons
equals [18]
\beq
f_d(p_{m,z},\vec{p}_{\perp})=
\left|\phi_{d}(p_{m,z},\vec{p}_{\perp})-
{\sigma_{tot}(pn) \over 16\pi^{2}}\int d^{2}\vec{k}\,
\phi_{d}(p_{m,z},\vec{p}_{\perp}-\vec{k})
\exp\left(-{1\over 2}B\vec{k}^{2}\right)
\right|^{2} \, .
\label{eq:3.2.1}
\endeq
Here $\phi_{d}(\vec{k})$ is the momentum-space wave function
of the deuteron.
The measured transparency factor $T_{d}$
depeds on the acceptance $p_{max}$:
\beq
T_{d}=
[\int^{p_{max}}d^{2}\vec{p}_{\perp}f_d(\vec{p}_{\perp})]
/[
\int^{p_{max}}d^{2}\vec{p}_{\perp}\phi_d(\vec{p}_{\perp})^{2}]
\label{eq:3.2.2}
\endeq
If $R_{d}^{2}p_{max}^{2} \gg 1$, but $Bp_{max}^{2} \ll 1$, which is
the case for the NE18, then
\beq
1-T_{d} \sim {\sigma_{tot}(pn) \over 2\pi R_{d}^{2}} \sim 0.07\,\, ,
\label{eq:3.2.3}
\endeq
which is about twice
the shadowing effect in the $\sigma_{tot}(Nd)$.
\noindent
\medskip\\
{\sl 3.3.  The theoretical interpretation of the NE18 data}
\medskip\\
The definition of nuclear transparency $T_{A}$ involves integration
of the measured and PWIA cross sections
over the experimental acceptance domain
$D$ in the
$(E_{m},p_{m,z},p_{\perp})$ space:
\beq
T_{A}=
{\int_{D} dE_{m}\, dp_{m,z}\,dp_{\perp}
S(E_{m},p_{m,z},p_{\perp}) \over
\int_{D} dE_{m}\, dp_{m,z}\,dp_{\perp}
S_{PWIA}(E_{m},p_{m,z}+\Delta p_{m,z},p_{\perp})}
\label{eq:3.3.1}
\endeq
The results of sections 3.1 and 3.2 show that the
measured spectral function
$S(E_{m},\vec{p}_{m})$
cannot be factorized into
$S_{PWIA}(E_{m},\vec{p}_{m})$
and an attenuation factor which is independent of the missing
energy and momentum.
The difference between $\sigma_{tot}(pN)$
and $\sigma_{in}(pN)$, and between $W^{(0)}$ and $T_{A}$ thereof,
is very large at moderate energies.
$T_{A}(NE18)$ includes
partly the elastically rescattered struck protons. Since the
diffraction slope B rises with proton energy,
the higher is $Q^{2}$ the larger is the fraction of the elastically
rescattered struck protons included in $T_{A}(NE18)$.
The effect of the shift (\ref{eq:3.1.4}) on $T_{A}$ depends on the
$p_{m,z}$-acceptance. It vanishes for the wide
$p_{m,z}$-acceptance. But
for the narrow acceptance centered
at $p_{m,z}=p^{*}$ it can be evaluated as
\beq
{T_{A}(\Delta p_{m,z}) \over T_{A}(\Delta p_{m,z}=0)}
\approx 1 +
{5\over 3}\cdot {(\Delta p_{m,z}+p^{*})^{2}-(p^{*})^{2}
 \over k_{F}^{2} }
\label{eq:3.3.2}
\endeq
and enhances $T_{A}$ by $\sim 3\%$ at $p^{*}=0$.
Here $k_{F}$ is the Fermi momentum.
More detailed calculations with the realistic
spectral functions are needed to improve upon the crude estimate
(\ref{eq:3.3.2}) [28].

In Fig.~5 we present our predictions [17,18] for $T_{A}$, $W^{(0)}$ and
$T_{A}(NE18)$ in which the $p_{\perp}$ integration for heavy nuclei is
extended up to $p_{max}=250\, {\rm MeV/c}$ as relevant to the NE18
situation [3].
We find very good quantitative agreement with the
NE18 data. The principle conclusion
from this comparison is that there is no signal of CT in the
$A(e,e'p)$ scattering at $Q^{2} \leq 7$ GeV$^{2}$.
CT effects start to set in at
$Q^{2}\sim 7$ GeV$^{2}$ for carbon with $T_{C}$ being increased by
$\sim 2\%$ at the largest $Q^{2}$ of the
NE18 experiment (Fig.~5b), for the
heavier nuclei the CT  effect is much smaller, see below.

Why even at $Q^{2}= 7$ GeV$^{2}$ the quark configurations emerging
from the $ep$ scattering vertex have large interaction cross section
so that
CT effect in $(e,e'p)$ is
not seen?
What makes the onset of CT in the vector meson production and
$(e,e'p)$ scattering so much different?
\noindent
\medskip\\
{\sl  3.4. It is not easy to shrink the ejectile state}
\medskip\\
The hard scattering process produces the ejectile state $|E\rangle$,
which then is projected onto the observed final state hadron.
In the exclusive production of vector mesons $\Psi_{E}(r)
\propto \sigma(r)\Psi_{\gamma^{*}}(r)$ [8]. Because of CT,
$\Psi_{E}(r)$ has a
hole at $r=0$, but at large $Q^{2}$ the
decrease of the wave function of the photon (\ref{eq:2.1.4})
takes over
and $\Psi_{E}(r)$ has a small size
$\sim 2 r_{Q}$. In terms of the ejectile state $|E\rangle$
the strength of FSI can be rewritten as
$\Sigma_{V}= \langle V|\sigma(r)|E\rangle/\langle V|E\rangle$.

In the quasielastic $A(e,e'p)$ scattering the energy $\nu$ and
$Q^{2}$ are tightly correlated, $2m_{p}\nu \approx Q^{2}$. The
moderately large $Q^{2}$ implies the moderately large $\nu$. The
transverse size $r$ of the ejectile can not be considered frozen.
The new formalism is needed. The problem can readily be treated in
the lightcone approach, but all the essential physics can be
presented in the framework of more familiar nonrelativistic
quantum mechanics, which we follow below.

In the lack of the frozen transverse size, it is more convenient
to work in the hadronic basis of the mass eigenstates $|i\rangle$
[5,6]. On the one hand, the ejectile state equals
\beq
|E\rangle = J_{em}(Q)|p\rangle=
\sum_{i}|i\rangle
\langle i|J_{em}(Q)|p\rangle =
\sum_{i}G_{ip}(Q)|i\rangle
\label{eq:3.4.1}
\endeq
where $G_{ip}(Q)$ are the $ep\rightarrow ei$ transition
form factors. The
strength of FSI will be given by ($G_{em}(Q)=G_{pp}(Q)$)
\beq
\Sigma_{ep}(Q) =  {\langle p|\hat{\sigma}|E\rangle \over
\langle p|E\rangle}=
{1 \over G_{em}(Q)}\sum_{i}\sigma_{pi}G_{ip}(Q)=\sigma_{tot}(pp)+
{1 \over G_{em}(Q)}\sum_{i\neq p}\sigma_{pi}G_{ip}(Q)
\label{eq:3.4.2} \, .
\endeq
Here $\hat{\sigma}$ is the cross
section or diffraction operator, which gives the forward
diffraction scattering amplitudes $f(jp\rightarrow kp)=
i\langle j|\hat{\sigma}|k\rangle = i\sigma_{jk}$.
The normalization is such that
$\sigma_{tot}(jN)=\sigma_{jj}$.
In the limit of weak FSI (cf. Eq.~(\ref{eq:2.1.6}))
\arr
T_{A} =
1-\Sigma_{ep}(Q){1\over 2A} \int d^{2}\vec{b}\,T(b)^{2}+...
\label{eq:3.4.3}
\endarr

On the other
hand, the ejectile wave function is found in any textbook in
quantum mechanics and/or nuclear/particle physics
(here
the $\vec{r}$-plane is normal to the momentum
transfer $\vec{Q}$):
\beq
\Psi_{E}(\vec{r},z)=
\exp\left({i\over 2}Qz\right)\Psi_{p}(\vec{r},z)\, .
\label{eq:3.4.4}
\endeq
Because
$|\Psi_{E}(\vec{\rho},z)|^{2}= |\Psi_{p}(\vec{\rho},z)|^{2}$,
this wave packet has the transverse size identical to the size of
the proton [4]! (The often made statement that the ejectile has a
small and/or even vanishing size is quite misleading.)
Which shows that starting with the such large-size
ejectile one will end up with weak FSI, {\sl i.e.}, with
small $\Sigma_{ep} \ll \sigma_{tot}(pp)$,
only provided that there is a special conspiracy
between the electromagnetic form factors $G_{ip}(Q)$ and the
diffraction scattering amplitudes $\sigma_{ip}$. What is an
origin of such a conspiracy?
\noindent
\medskip\\
{\sl 3.5.   Color transparency sum rules and vanishing FSI: conspiracy
of hard and soft scattering}
\medskip\\
In QCD the (anti)quarks in the hadron always have the one-gluon
exchange Coulomb interaction at short distances. This Coulomb
interaction leads to a very special asymptotics of the
form factor at $Q^{2} >> R_{p}^{2}$ [29] (for the sake of simplicity
we discuss the two-quark states)
\beq
G_{ik}(\vec{Q})
\propto {V(\vec{Q}) \over \vec{Q}^{2}}\Psi_{i}^{*}(0)\Psi_{k}(0)
\label{eq:3.5.1}
\endeq
Here $\Psi_{i}(0)$ are the wave functions at the origin, and
$V(Q)$ is the one-gluon exchange quark-quark hard
scattering amplitude.
Making use of the QCD asymptotics (\ref{eq:3.5.1}) we find
\arr
\Sigma_{ep}(Q) = {1 \over G_{em}(Q)}
\sum_{i}\sigma_{pi}G_{ip}(Q)
\propto {1\over G_{em}(Q)}{V(\vec{Q}) \over \vec{Q}^{2}}
\sum_{i}\sigma_{pi}\Psi_{i}^{*}(0)\Psi_{p}(0)
\propto \sum_{i}\sigma_{pi}\Psi_{i}^{*}(0)
\label{eq:3.5.2}
\endarr
Remarkably, the {\sl r.h.s.} of Eq.~(\ref{eq:3.5.2})
vanishes by virtue of CT sum
rules [5,6]. The proof goes as follows:
The state of transverse size $\vec{r}$ can
be expanded in the hadronic basis as
$
\left.|\vec{r}\right>=
\sum_{i}\Psi_{i}(\vec{r})^{*}\left.|i\right> $
and the hadronic-basis expansion for $\sigma(r)$ reads  as
\beq
\sigma(\rho)=\langle \vec{r}|\hat{\sigma}|\vec{r}\rangle =
\sum_{i,k}
\left<\vec{r}|k\right>
\left<k|\hat{\sigma}|i\right>\left<i|\vec{r}\right>=
\sum_{i,k} \Psi^{*}_{i}(\vec{r})\Psi_{k}(\vec{r})
\sigma_{ki}
\label{eq:3.5.3}
\endeq
By virtue of CT $\sigma(r=0)=0$, and we obtain
the "CT sum rule" [5,6,9]
\beq
\sum_{i,k} \Psi_{k}(0)^{*}\Psi_{i}(0)\sigma_{ki} = 0      \, .
\label{eq:3.5.4}
\endeq
Considering the matrix elements $\langle \vec{r}|\hat{\sigma}|k\rangle$
at $\vec{r} \rightarrow 0$, one readily finds a whole family of CT
sum rules [6,19]
\beq
\sum_{i} \Psi_{i}(0)\sigma_{ik} = 0      \, .
\label{eq:3.5.5}
\endeq
It is precisely the sum rule (\ref{eq:3.5.5}), which in conjunction
with the QCD asymptotics of the electromagnetic form factors
(\ref{eq:3.1.1}) ensures that
the strength of FSI (\ref{eq:3.4.2},\ref{eq:3.5.2})
vanishes. Above we have implicitly assumed
the asymptotically large $Q^{2}$, so
that all the excited states contribute the CT sum rules.
Evidently, the onset of CT is controlled by how rapidly the
CT sum rules are saturated in the truncated basis of few states
which can contribute at finite $Q^{2}$. This saturation is
controlled by the coherency constraint [5,6,19].
\noindent
\medskip\\
{\sl 3.6.   The onset of CT and the coherency constraint}
\medskip\\
Hitherto the operator $\hat{\sigma}J_{em}$ in
the definition of $\Sigma_{ep}$ was treated
as a local operator. By the
nature of CT experiments the ejectile state produced on one
nucleon is probed by its interaction with other nucleons
a distance $\Delta z \sim R_{A}$ apart.
{}From the kinematics
of deep inelastic scattering
$
m_{i}^{2}-m_{p}^{2}=2m_{p}\nu - Q^{2}-2\nu k_{z}$,
and  different components $|i\rangle$
of the ejectile wave packet are produced
with the longitudinal momenta differing by [5]
\beq
\kappa_{ip}=m_{p}{ m_{i}^{2} -m_{p}^{2} \over Q^{2}} =
{ m_{i}^{2} -m_{p}^{2} \over 2\nu }       \, .
\label{eq.3.6.2}
\endeq
The different components of the ejectile state develop the
relative phase $\kappa_{ip}\Delta z$ from the production to the
rescattering point, and more accurate evaluation of $\Sigma_{ep}$
gives  [5,6,18,19]
\arr
\langle p|\hat{\sigma}J_{em}|p\rangle =
\sum_{i}\langle p|\hat{\sigma}|i\rangle\langle i|J_{em}|p\rangle
{}~~~~~~~~~~~~~~~~~~~~~~~~~~~
\nonumber\\
\Longrightarrow
\sum_{i}\langle p|\hat{\sigma}|i\rangle\langle i|J_{em}|p\rangle
\exp[i\kappa_{ip}(z_{2}-z_{1})]
\Longrightarrow
\sum_{i}\langle p|\hat{\sigma}|i\rangle\langle i|J_{em}|p\rangle
G_{A}(\kappa_{ip})^{2}\,\, ,
\label{eq:3.6.3}
\endarr
Here $G_{A}(\kappa)$ is the body (charge) formfactor of the nucleus
which emerges after averaging of the phase factor
$\exp[i\kappa (z_{2}-z_{1})]$ over positions of the production
and rescattering points inside the nucleus.
The finite-energy expansion for the strength of FSI takes the form
[5,6,18,19]
\beq
\Sigma_{ep}(Q)=\sigma_{tot}(pN)+
\sum_{i\neq p}{G_{ip}(Q) \over G_{em}(Q)}\sigma_{ip}G_{A}(\kappa_{ip})^{2}
\label{eq:3.6.4}
\endeq
In the low-energy limit of $\kappa_{ip} > R_{A}$, {\sl i.e.},
\beq
Q^{2} \lsim 2R_{A}m_{p}\Delta m \sim 5\cdot A^{1/3} {\rm GeV}^{2}
\label{eq:3.6.5}
\endeq
 all inelastic
channels decouple,
$\Sigma_{ep}=\sigma_{tot}(pN)$, and FSI measures the free-nucleon
cross section.

Eq.~(\ref{eq:3.6.4}) answers all questions on the
onset of CT: i) Only those intermediate hadronic states which
are diffractively produced in the forward $pN$ scattering
do contribute to the CT sume rule and to the CT reduction of
the strength of FSI. The $N^{*}(1680)$ is a prominent channel [30],
which suggests $\Delta m \sim 0.7$ GeV and the threshold for CT is
$Q^{2} > 5\cdot A^{1/3} {\rm GeV}^{2}$.
ii) The diffraction
dissociation amplituides are small, and opening of few channels
leads to a little reduction of $\Sigma_{ep}$. For instance, for
the prominent excitation of $N^{*}(1680)$ the cross section is
$\approx 0.2$ mb compared to the elastic scattering of $\approx 7$
mb. Opening of this channel could produce only $\sim 15\%$ signal
of CT.
iii) Still another prominent channel is an excitation of the
continuum $\pi N$ states (the Drell-Hiida-Deck process [29])
with about the same cross section. The mass spectrum of this
continuum is peaked at $\sim 1.4$ GeV. Its effect on $\Sigma_{ep}$
can be interpreted as stripping of the pionic cloud off the nucleon.
The bare nucleon interacts with somewhat smaller cross section,
which also contributes $\sim 15\%$ to the signal of CT.
\pagebreak\\
\noindent
{\sl 3.7. The scaling behavior of nuclear
transparency for $A(e,e'p)$}
\medskip\\
In Fig.6   we show how $\Sigma_{ep}$ decreases with $Q^{2}$ in
the realistic QCD inspired model [5,6,19] of the diffraction operator
$\hat{\sigma}$, which also incorporates excitation of the $\pi N$
continuum states. The CT effect in $T_{A}(Q^{2})$ follows closely
the variation of $\Sigma_{ep}(Q^{2})$.
 The values of
$Q^{2}$ studied by NE18 correspond to the threshold of CT for
the carbon nucleus.
We predict [5,19] a substantial
CT effect at $Q^{2} \sim 20$ GeV$^{2}$, which are accessible at
ELFE (European Laboratory for Electrons) and SLAC, and could
well be observed at SLAC should the NE18 experiment be upgraded.

It is instructive to look at this slow onset of CT in a
somewhat different language. The coherency constraint embodied in the
nuclear form factor
$G_{A}(\kappa_{ik})$ eliminates the high-mass components of the
ejectile wave packet, and the effective ejectile state can be
written as
\beq
|E_{eff}\rangle =\sum_{i}G_{ip}(Q)G_{A}(\kappa_{ip})|i\rangle
\propto \psi_{p}(0)\sum_{i}\Psi_{i}^{*}(0)\Psi_{i}(\vec{r})
G_{A}(\kappa_{ip}) \, ,
\label{eq:3.7.1}
\endeq
which can be compared with the delta-function
$\delta(\vec{r})=\sum_{i}\Psi_{i}^{*}(0)\Psi_{i}(\vec{r})$.
The true ejectile state Eq.~(\ref{eq:3.4.1}) as probed
at the production vertex has {\sl exactly the same} transverse size
as the free proton.
Asking for the quasielastic scattering, one projects this
ejectile state onto the final state proton. This projection,
in concert with the coherency condition, imposes a constraint
on the subset of $N_{eff}$ frozen out intermediate states
for which $G_{A}(\kappa_{ip}) \sim 1$ and which
could contribute coherently
to the intranuclear FSI and CT effect.

The states $|i\rangle$ with the mass
\beq
m_{i}^{2} \lsim M^{2} = {Q^{2}\over R_{ch}(A)m_{p}}
\label{eq:3.7.2}
\endeq
 contribute to the
the truncated wave packet $|E_{eff}\rangle =\sum_{i}^{N_{eff}}
\Psi_{i}^{*}(0)\Psi_{i}(\vec{r})$.
Naively, one would have expected
that the least possible size of such a wave packet satisfies
$r^{2}_{min}\propto 1/M^{2} \propto  1/Q^{2}$, but in practice the
decrease of $r^{2}_{min}$ with $Q^{2}$ is much slower.
For instance, in the harmonic oscillator models [5,19]
\beq
r^{2}_{min} \propto \left({R_{ch}(A)m_{p}\over Q^{2}  }\right)^{\eta}
\label{eq:3.7.3}
\endeq
with the (model-dependent) exponent $\eta \sim    0.2$.
Evidently, $r^{2}_{min}$ only depends on $M^{2}$ and the
specific dependence on $R_{A}$ in the {\sl r.h.s.} of
Eq.~(\ref{eq:3.7.3}) comes from the coherency constraint
Eq.~(\ref{eq:3.7.2}).
In the regime of CT we find the scaling law (cf. Eq.~(\ref{eq:2.6.1}))
\beq
1-T_{A}(e,e'p) \propto {A \over R_{ch}(A)^{2}}r^{2}_{min} \sim
{A \over R_{ch}(A)^{2}}\cdot
\left({R_{ch}(A)m_{p}\over Q^{2}  }\right)^{\eta} \,\,.
\label{eq:3.7.4}
\endeq
which is quite different from the one proposed by
Ralston and Pire [31], who did not take into account
the coherency constraint.

Now we are in the position to tell the difference between the
E665 and NE18 experiments:
In the E665 energy $\nu$ was very high, $2m_{p}\nu \gg  Q^{2}$
and all intermediate states do contribute without hitting the
coherency constraint. Futhermore, small $r$ is taken care of
by the wave function of the photon (\ref{eq:2.1.4}) which
guarantees small size of the ejectile. In the NE18 experiment
$2m_{p}\nu = Q^{2}$ by
the kinematics of the quasielastic $(e,e'p)$ scattering, the
energy is not high and the coherency constaint drastically
delays the onset of CT. Furthermore, here one starts with the
large-size ejectile Eq.~(\ref{eq:3.4.4}).
\noindent
\medskip\\
{\sl 3.8. CT can be seen at CEBAF}
\medskip\\
In spite of the (theoretically anticipated [5,6])
failure to observe  CT effect in the SLAC
NE18 experiment, one must not overlook the potential of the
dedicated $^{4}He(e,e'p)$ scattering experiment at CEBAF.
We specifically suggest looking at the sruck protons with
$p_{\perp} > k_{F}$ [18]. The yield of such protons will be
proportional to the elastic rescattering probability $W_{1}$
Eq.~(\ref{eq:3.1.6}). In the regime of CT [6,18,19]
\beq
W_{1} \propto  \eta^{2}=
\left({\Sigma_{ep}(Q^{2})\over \sigma_{tot}(pN)}\right)^{2}
\label{eq:3.8.1}
\endeq
With allowance for the center-of-mass motion, for the light
nuclei [18]
\beq
\Sigma_{ep}(Q^{2})=\sigma_{tot}(pN)+\sum_{i\neq p}\sigma_{ip}
{G_{ip}(Q^{2})\over G_{pp}(Q^{2})}G_{A}({2A\over A-1}\kappa_{ip}^{2})
\, .
\label{eq:3.8.2}
\endeq
Our prediction for the strength of the elastic rescattering $\eta^{2}$
for $^{4}He(e,e'p)$ reaction is shown in Fig.7.
Because of the small size of the $^{4}He$ nucleus, we predict
a rapid onset of CT in $^{4}He(e,e'p)$ scattering, which can be
detected in the dedicated CEBAF measurements of the
elastic-rescattering tail of the $p_{\perp}$ distribution.
\noindent
\medskip\\
{\sl 3.9.   On the accuracy of the uncorrelated
Glauber model}
\medskip\\
The above evaluations of
FSI effects in $A(e,e'p)$ scattering were mostly based on the
independent-particle model for nuclear wave functions. Recently there
was much discussion of the short-range correlation effects for $T_{A}$
[31,15]. The principle conclusion is that the correlation effect is
diluted by the finite radius of $pN$ interaction and further
reduced by cancellations of the hole [31] and spectator [15]
effects, so that
the uncorrelated Glauber model calculations of nuclear transparency
$T_{A}$ are good to a few per cent accuracy [15].
\noindent
\bigskip\\
{\bf 4.~ Conclusions}
\medskip\\
The E665 observation [8] of the onset of CT is a major breakthrough
in the subject of CT and paves the way for dedicated experiments
on CT. The E665 effect was predicted [11] (not postdicted!), and
the agreement betwen the theory and experiment is very good. The
nonobservation of CT in the NE18 experiment [3] is also
a good confirmation of theory [5]. We have a good understanding of
why the onset of CT in $A(e,e'p)$ and $\gamma^{*}A\rightarrow
VA,\,VA^{*}$
reactions is so strikingly different. In a way, the mechanism of
CT in $A(e,e'p)$ scattering is much subtler, and the potential of
CT experiments at CEBAF must not be overlooked.
\noindent
\medskip\\
{\bf 5.~ Acknowledgements}
\medskip\\
One of the authors (N.N.N) is grateful to O.Benhar, A.Fabrocini and
S.Fantoni for organizing this most stimulating conference.
Thanks are due to B.Fillipone, A.Lung, R.McKeown and R.Milner
for discussions
and communications on the NE18 experiment, and to G.Fang and
H.Schellman for discussions on the E665 experiment.
We are grateful to J.Nemchik and A.Szczurek for their assistance.

{\bf \LARGE Figure captions:}

\begin{itemize}

\item[Fig.~1 - ]
Predictions [11] of nuclear transparency $T_{A}=\sigma_{A}/A\sigma_{N}$
for the incoherent
exclusive production of $\rho^{0}$ mesons vs. the E665 data [14].

\item[Fig.~2 - ]
Predictions [11] of nuclear transparency
$T_{A}^{(coh)}/T_{C}^{(coh)}
=[144d\sigma_{A}/A^{2}d\sigma_{C}]_{\vec{q}\,^{2}=0}$
for the forward coherent production of the $\rho^{0}$ mesons.

\item[Fig.~3 - ]
Predictions [11] of the $Q^{2}$ dependence of the
ratio of cross sections
$R_{coh}(A/C)=
12\sigma_{A}/A\sigma_{C}$
for coherent production of the $\rho^{0}$ mesons vs. the E665
data [14]. The arrows indicate predictions for the complete CT
at $Q^{2}\rightarrow \infty$.

\item[Fig.~4 - ]
Predictions [11] of the $Q^{2}$ dependence of exponents of
parametrizations $\sigma_{A}(inc) \propto A^{\alpha_{inc}}$,~~
$[d\sigma_{A}(coh)/dq^{2}]_{\vec{q}\,^{2}=0} \propto
A^{\alpha_{coh}(\vec{q}\,^{2}=0)}$ and $\sigma_{A}(coh) \propto
A^{\alpha_{coh}}$
vs. the E665
data [14].

\item[Fig.~5 - ]
   Predictions for the $Q^{2}$ dependence of nuclear transparency:
   the dashed curve denotes the $p_{\perp}$-integrated nuclear
   transparency $T_{A}$; the dotted curve gives the nuclear transparency
   $W^{(0)}$ measured in parallel kinematics at $p_{\perp}=0$;
   the solid curve represents the nuclear transparency $T_{A}(NE18)$
   including the NE18 acceptance cuts [2]; the
   dot-dashed curve in panel b)  shows the effect of the onset of CT.

\item[Fig.~6 - ]
Predictions [19] for the $Q^{2}$-dependence of the strength of
FSI $\Sigma_{ep}$
and nuclear transparency $T_{A}$.
 \item[Fig.~7 - ]
   Predictions [18] for the CT effect in the rate of
   $^{4}He(e,e'p)$ reaction with production of high transverse
   momentum protons. The quantity shown is the relative
   strength of the elastic-rescattering tail in the
   $p_{\perp}$-distribution which is given by $\eta^{2}=
   [\Sigma_{ep}(Q^{2})/\sigma_{tot}(pN)]^{2}$

\end{itemize}
\end{document}